\RequirePackage{fix-cm}
\documentclass{article} 
\usepackage{graphicx}

\usepackage{hyperref,xcolor,soul,amsmath}
\usepackage[colorinlistoftodos]{todonotes}
\usepackage{authblk}
\usepackage{pdfpages}
\title{Adapting a plant tissue model to animal development: introducing cell sliding into {\em VirtualLeaf}
\footnote{This work was cofinanced by the Netherlands Consortium for Systems Biology (NCSB; 2008-2013), which was part of the Netherlands Genomics Initiative/Netherlands Organisation for Scientific Research (HBW and RMHM). This work was also part of the research program ‘‘Innovational Research Incentives Scheme Vidi Cross-divisional 2010 ALW’’ with project number 864.10.009 to RMHM, which is (partly) financed by the Netherlands Organization for Scientific Research (NWO). The simulations were carried out on the Dutch national e-infrastructure with the support of SURF Cooperative (www.surfsara.nl).
Support for LAD and partial support for HBW were provided by grants from the National Institutes of Health (NIH R01 HD044750 and R21 ES019259) and the National Science Foundation  (NSF CMMI-1100515). Any opinions, findings, and conclusions or recommendations expressed in this material are those of the authors and do not necessarily reflect the views of the NSF or the NIH.
}
}

\author[1,2,4]{Henri B. Wolff}
\author[2]{Lance A. Davidson\footnote{Co-corresponding authors: {\tt lad43@pitt.edu} and {\tt merksrmh@math.leidenuniv.nl}}}
\newcommand\CoAuthorMark{\footnotemark[\arabic{footnote}]} 
\author[1,3,5]{Roeland M. H. Merks\protect\CoAuthorMark}
\affil[1]{Centrum Wiskunde \& Informatica, Science Park 123, 1098 XG Amsterdam, the Netherlands}
\affil[2]{Department of Bioengineering, Bioscience Tower 3-5059, 3501 Fifth Avenue, University of Pittsburgh, Pittsburgh, PA, USA}
\affil[3]{Mathematical Institute, University Leiden,  P.O. Box 9512, 2300 RA Leiden,  the Netherlands}
\affil[4]{Present address: Decision Modeling Center VUmc,  Department of Epidemiology and Biostatistics, VU University Medical Center, PO Box 7057, 1007 MB Amsterdam, the Netherlands}
\affil[5]{Present address:  Mathematical Institute and Institute of Biology, University Leiden,  P.O. Box 9505, 2300 RA Leiden,  the Netherlands}

\date{\today}
\begin{document}
\maketitle

\begin{abstract}
Cell-based, mathematical modeling of collective cell behavior has become a prominent tool in developmental biology. Cell-based models represent individual cells as single particles or as sets of interconnected particles, and predict the collective cell behavior that follows from a set of interaction rules. In particular, vertex-based models are a popular tool for studying the mechanics of confluent, epithelial cell layers. They represent the junctions between three (or sometimes more) cells in confluent tissues as point particles, connected using structural elements that represent the cell boundaries. A disadvantage of these models is that cell-cell interfaces are represented as straight lines. This is a suitable simplification for epithelial tissues, where the interfaces are typically under tension, but this simplification may not be appropriate for mesenchymal tissues or tissues that are under compression, such that the cell-cell boundaries can buckle. In this paper we introduce a variant of VMs in which this and two other limitations of VMs have been resolved. The new model can also be seen as on off-the-lattice generalization of the Cellular Potts Model. It is an extension of the open-source package VirtualLeaf, which was initially developed to simulate plant tissue morphogenesis where cells do not move relative to one another.  The present extension of VirtualLeaf introduces a new rule for cell-cell shear or sliding, from which T1 and T2 transitions emerge naturally, allowing application of VirtualLeaf to problems of animal development. We show that the updated VirtualLeaf yields different results than the traditional vertex-based models for differential-adhesion-driven cell sorting and for the neighborhood topology of soft cellular networks.

\end{abstract}

\section{Introduction}
\label{sec:intro}
How cells form tissues, organs, and organisms remains one of the most intriguing and most central questions of biology. Recent theoretical approaches to study collective cell behavior are taking a prominent role in addressing these questions. Theoretical approaches provide deeper intuition about processes that typically are unfamiliar to the researchers by testing the physical plausibility of a speculative hypotheses or by making predictions that can be tested experimentally. Theory can aid the analysis of data-rich time-lapse images of cell movements during development \cite{Brodland:2014cb,Merkel:2016de} by simulating how the behavior of individual cells might lead to collective behavior and how collective movements might influence individual cell behaviors. To support theoretical analysis of tissue formation, a large range of mathematical methods have been proposed. These range from systems of partial differential equations models (see, e.g,. Ref.~\cite{Keller1970vv,Keller.jtb71,Painter:2015cj}) to discrete methods that describe dense multicellular structures as interacting particle systems (see, e.g.,  \cite{JAGlazier:1993uk,Graner:1992ve,Newman:2005ux,Sozinova:2006fk,VossBohme:2010gf,Woods:2014bf,Smeets:2016eb,Barton:2017bv,Ghaffarizadeh:2018de} and Ref.~\cite{Liedekerke:2015de} for review). 

In contrast to continuum models, such so called cell-based \cite{RMHMerks:2005kw,Merks:2015cc}, or single-cell-based methods \cite{Anderson-Book2007} have the disadvantage that formal dynamical analyses are impossible except in relatively simple cases \cite{VossBohme:2010gf}. Nevertheless in many biological applications cell-based models are preferred as they can  incorporate `biological-rules' that reflect more physiologically realistic biology than can be achieved easily in continuum methods. For example, Odell et al. (1981) posed a calcium sensitive feedback system to spread a contraction wave \cite{Odell:1981db}.  As developmental biologists adopt biophysical methods and borrow principles of control theory to explain tissue formation, simulations will need to capture  interactions between multiple cell types and the diverse forms of cell-cell communication those interactions encode \cite{Lander:2007be}. Models will further need to integrate structural, mechanical, and biochemical cues with downstream effectors of morphogenesis such as cell division, cell death events,  and cell differentiation events \cite{Maree:2002hd,Hester:2011cc,Boas:2015td,Palm:2016jl}. Cell-based simulation methods provide a rich framework to study multiscale phenomena such as these, in that they simulate the dynamics of the cell and the tissue as a whole, while subcellular dynamics can be naturally integrated, such as gene regulation, secretion of signaling molecules,  the dynamics of the cytoskeleton, and electrophysiological mechanisms \cite{Boas:2014jd,Belmonte:2016bd,Sluka:2016cl,Kudryashova:2017gb}. Thus cell-based modeling approaches enable integration of the physics of collective cell behavior with diverse modes of subcellular biological regulation.

A large range of cell-based modeling techniques are available; they can be roughly classified into single-particle and multi-particle methods, and lattice-based and off-lattice techniques \cite{Merks:2015cc}. Single-particle techniques are efficient computationally and have found wide application, but they also have limitations.  For example, cells shape can affect the outcome of cell-cell interactions: cells that mutually attract one another via a chemoattractant form network-like structures if they are elongated, while they from separate `islands' if they are rounded \cite{RMHMerks:2006jp}.  Also, it can be important that subcellular compartments interact with their local environment relatively independently from one another. For example, contact inhibition of cellular protrusions can promote directional migration of neural crest cells  \cite{CarmonaFontaine:2008ft}. Although it is possible to simulate such problems using single-particle-based methods (see, e.g., Ref.~\cite{Palachanis:2015bea} for effects of cell shape in vascular patterning, Ref.~\cite{Woods:2014bf} for neural crest cell migration; and for cell sorting see, e.g., Refs.~\cite{Sulsky:1984uw,Graner:1993tg,Osborne:2017ff}), multiparticle methods allow for a large flexibility in cell shape that is more directly related to cell shapes acquired from time-lapse microscopy. Multiparticle methods are also better suited for the simulation of local mechanisms responsible for collective behavior (e.g., contact inhibition \cite{Woods:2014bf}),  as they make it possible to reduce cell-level assumptions to subcellular mechanisms.

Two widely used multiparticle techniques for cell-based modeling include the Cellular Potts Model (CPM) and the Vertex-based model (VM). The CPM represents cells as (usually connected) domains of lattice sites on a regular lattice. Cells move on the lattice by randomly extending or retracting their domain to adjacent lattice sites, according to a Hamiltonian energy function that describes the contractile  and viscoelastic structures that form each cell, the physical adhesive interactions between cells and, in some cases, extracellular materials. Alternatively,  structural elements have been used to describe the cell boundaries and cross-linking elements for the cell interior \cite{Odell:1981db}. A simplified version of this model are the vertex-based models (VM) \cite{Weliky:1990wn,Honda:1983kn,Staple:2010bn}. These represent the junctions between three (or sometimes more) cells in confluent tissues as point particles, connected using structural elements that represent the cell boundaries. Where the CPM defines tissues as assemblages of cells with individual cells represented as collections of adjacent lattice sites, VM describe the tissue as a polygonal tessellation of junctionally connected cells with each cell represented by a series of nodes representing three-cell junctions.    

In principle, CPM and VM are equivalent.  Like cells in CPMs, the dynamic movements of cells in VMs are driven by the physical properties of cell-cell interfaces, which are governed by a Hamiltonian function that usually includes interfacial tensions, cell adhesion,  and cell area constraints.  The parameters of CPM or VM can often be rescaled such that the model can be run in the other formalism \cite{Magno:2015bw}. However, some cases can arise where the two modeling frameworks cannot be interconverted \cite{Osborne:2017ff}. For instance, because the string-like elastic elements in their basic formulation describe cell-cell interfaces and require all cells to be interconnected, VMs are unsuitable for non-confluent tissues. In this paper we discuss further limitations of traditional VMs including: (a) description of cell-cell interfaces as straight lines,  (b) separation of membrane fluctuations and model dynamics, and (c) algorithms that represent cell neighbor changes with explicit topological transformations (e.g. so called T1 and T2 transitions).

In this paper we introduce a variant of the VM in which these three limitations have been resolved. The model is an extension of the open-source package VirtualLeaf~\cite{RMHMerks:2011cj,Merks:2012fu} (also see the re-engineered derivative {\em Virtual Plant Tissue}~\cite{DeVos:2017hn}), which was initially developed to simulate plant tissue morphogenesis where cells do not move relative to one another. VirtualLeaf differs from traditional VMs in that (a) cell interfaces are represented by multiple nodes that allow membrane fluctuations; (b) tissue topology changes exclusively through cell division with no T1 or T2 transitions; and (c) tissue dynamics are advanced used a Metropolis algorithm that incorporates membrane fluctuations. The present extension of VirtualLeaf introduces a new rule for cell-cell shear or sliding, from which T1 and T2 transitions emerge naturally, allowing application of VirtualLeaf to problems of animal development.

We will discuss two cases for which the updated VirtualLeaf yields different results than traditional VM. First, we discuss simulations of differential adhesion driven cell sorting, and show that the new update rule for cell sliding facilitates complete cell sorting. We will then turn our attention to epithelial dynamics, and discuss cases for which the flexibility of cell membranes affects the neighborhood topology of soft networks \cite{Farhadifar:2007hr}.

\section{Methods}
\begin{figure}
\includegraphics[width=\textwidth]{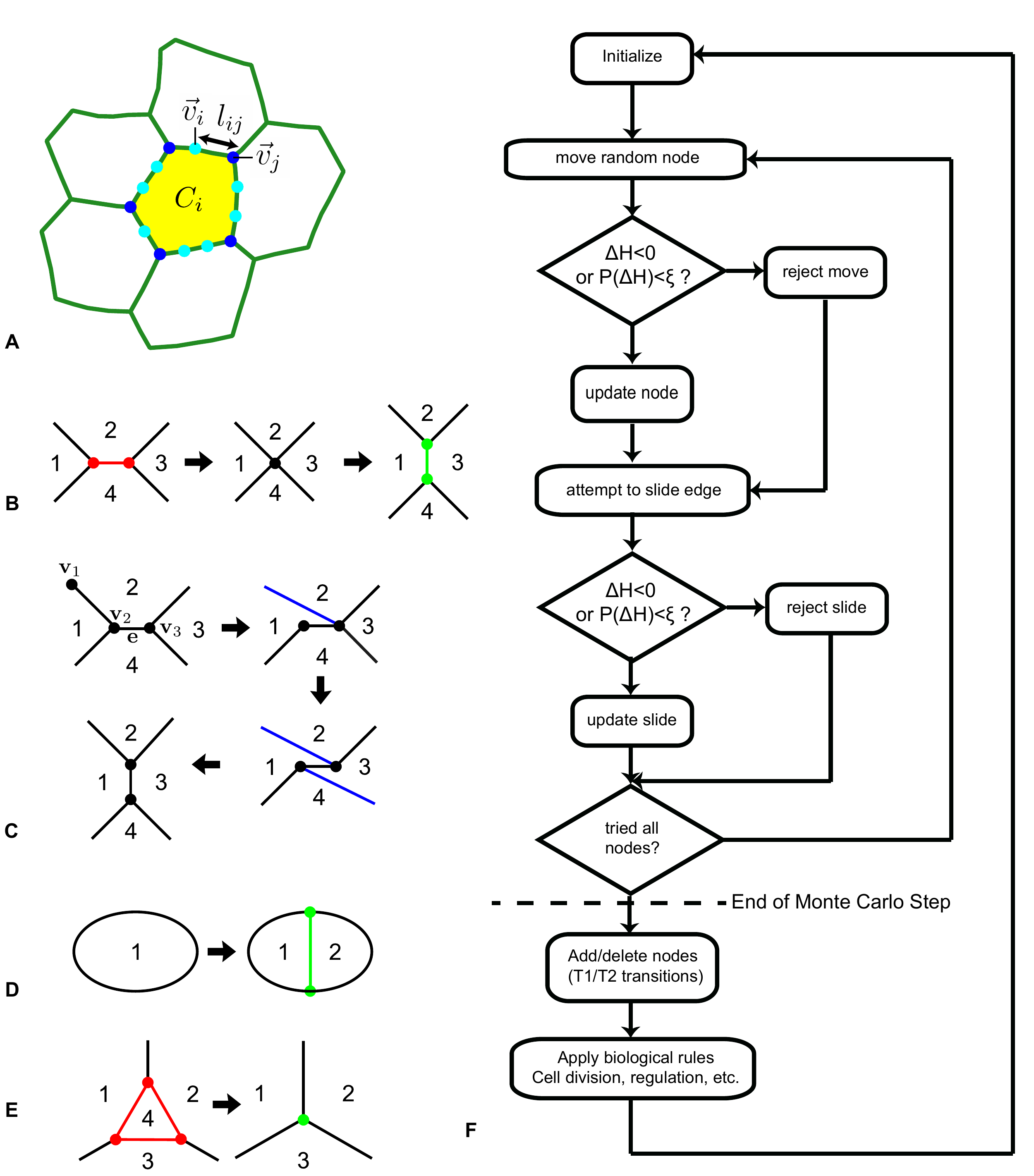}
\caption{{\bf Overview of the cell-based model}
(A) Polygonal representation of a collection of cells. Cell $C_i$ consists of edges (green) $l_{ij}$ connected by nodes $\vec{v}_i$  and $\vec{v}_j$. Nodes that connect three or more cells are shown in blue. The 2-connected nodes (shown in blue) account for membrane flexibility.  (B-E) Topological rearrangements of vertices and edges. Numbers represent cells. New vertices and edges are green, and red vertices and edges are to be removed. Blue edges moved by sliding. (B) Traditional approach through T1 transitions: one edge is added and one edge is removed; (C) Novel approach through slide events having the same topological effect as the T1 transition shown in panel B; (D) Cell division; (E) T2 transition: a cell is removed from the tissue and replaced with a 3-connected node; (F) Flow chart of an extended VirtualLeaf simulation. During a Monte Carlo step VirtualLeaf attempts to move and slide all nodes once in a random order. After one such loop the network is rearranged, and ``biological rules'' are applied }
\label{fig:fig1}
\end{figure}

VirtualLeaf represents confluent tissues in two dimensions as a set of interconnected polygonal cells.  A cell $C_i=\{V_i, E_i, \alpha_i\}$ is defined by a set of $n$ vertices, $V_i=\{\vec{v}_1\dots \vec{v}_n\}$ that are connected by $m$ edges, $E_i=\{\vec{e}_1 \ldots \vec{e}_m\}$, and a set of cell attributes, $\alpha_i$. Adjacent cells share the same vertices and edges. Thus, the tissue $T=\{C,V,E\}$ is defined by the set of all cells in the tissue, $C$, and by all vertices  in the tissue, $V=\bigcup_{i\in T}V_i$ and all edges, $E=\bigcup_{i\in T}E_i$  (Fig~\ref{fig:fig1}A). A Hamiltonian function, $H$, describes the balance of passive, mechanical forces in the tissue, including, adhesive forces between cells, membrane tensions and expansive cellular forces. The exact form of the Hamiltonian differs between models; in its simplest form \cite{RoelandMHMerks:2011cj} it includes a volume conservation term to resist compression of the cells and a line tension term to resist expansion of the membranes,
\begin{equation}
H=\lambda_A\sum_{c\in C}\left(A(c)-A_T(c)\right)^2+\lambda_M\sum_{\vec{e}\in E}\left(\|\vec{e}\|-L_T\right)^2.
\label{eq:H}
\end{equation}
The first term on the right-hand side (RHS) is the volume conservation term. Here $A_T(c)\in\alpha_c$ is the resting area of cell $c$, \textit{i.e.}, the area it would take up in absence of counteracting compressive or expansive forces,  $A(c)$, the actual area of cell $c$, and $\lambda_A$ is a Lagrange multiplier. The second RHS term gives the energy of the cell boundary, which is represented as a set of connected springs of rest length $L_T$. The sum runs over the edges $e$ taken from $E$, the set of all edges in the simulation, and $\lambda_M$ is a Langrange multiplier. 

We update the model using Metropolis dynamics: we iteratively select a random node  $\vec{v}_i$ and attempt to move it to a randomly chosen new position $\vec{v}_i\prime=\vec{v}_i+\vec{\xi}\Delta x$, with $\vec{\xi}\in\{[-1/2,1/2],[-1/2,1/2]\}$, i.e., a random vector chosen uniformly from a square of size $1\times1$ centered at $(0,0)$, and $\Delta x$ the step size. The algorithm calculates the change of the Hamiltonian resulting from the attempt, $\Delta H$, and accepts the move if $\Delta H<0$. To keep the system from settling into local minima, and to mimic active, random cell motility, we also accept moves increasing the Hamiltonian $\Delta H>0$ with Boltzmann probability $P(\Delta H)=\exp(-\Delta H/T)$, with the Boltzmann temperature, $T$, setting the amount of random cell motility or `noise' added in this way.  

The key novelty that makes the model applicable to animal tissues,  is that here we allow cells to move through the tissue. To this end, we introduce a {\em sliding operator} to further reduce the Hamiltonian (Figure~\ref{fig:fig1}C). The sliding operator allows an edge $\vec{e}$, connecting nodes $\vec{v}_1$ and a node $\vec{v}_2$ that must be connected to 3 or more nodes, to `hop' from $\vec{v}_2$ to a third node $\vec{v}_3$ connected to $\vec{v}_2$. Similar to the regular moves, a slide is accepted with probability $P(\Delta H)=\{1,\Delta H<0;\exp(-\Delta H/T),\Delta H>0\}$. 

During one Monte Carlo Step (MCS), we cycle over all nodes $v$ in random order. For each node, we first attempt to move it. If the node is of order 3 or higher, we also try to slide it (see flowchart in Figure~\ref{fig:fig1}F). After completion of one MCS,  the descriptions of the cell membranes are refined if necessary, so as to keep an approximately even distribution of edge lengths. To do so, all edges $\vec{e}\in E$ whose length exceeds a threshold, $\|\vec{e}\|>l_\mathrm{max}$, are split into two by inserting a new vertex in the middle. Similarly, if $\|\vec{e}\|<l_\mathrm{min}$, the edge is deleted from the tissue, replacing the two vertices and their connections for a new, fused vertex containing the connections of the two original vertices combined. 

Independent of this Hamiltonian description of cell mechanics and cell motility, additional rules motivated by the biological problem can be included, including cell growth, cell division, and subcellular models describing the genetic or metabolic networks regulating cell behavior using differential equations \cite{RoelandMHMerks:2011cj}.  If the additional rules can be safely assumed to run at a much slower rate than the cellular mechanics, we make a quasi-steady state assumption for the cellular mechanics: First, we iterate the Metropolis dynamics until the Hamiltonian has practically stabilized, that is if $\Delta H/\Delta t<\epsilon$, with $\epsilon$ a small number; then, we apply the additional rules for a number of time steps. In other models (i.e. the cell sorting model below) the Metropolis algorithm describes a kinetic mechanism that does not stabilize within the course of a simulation. In those cases, we apply an “operator splitting approach” in which the Monte Carlo steps are alternated with steps of the additional rules.  

\section{Results}
\label{sec:results}
We validate the model extensions by looking at two classical problems: (a) Differential adhesion cell sorting \cite{JAGlazier:1993uk,Graner:1992ve} and (b) cell packing in epithelial monolayers \cite{Farhadifar:2007hr}. VirtualLeaf provides new insight into both problems.

\subsection{Cell sorting}
\label{sec:cellsorting}
Classic experiments by Holtfreter (reviewed in Ref.~\cite{Steinberg:1996cd}) have shown that cells of different embryonic tissues can phase-separate. A number of (strongly related) hypotheses have been proposed to explain this phenomenon. Steinberg \cite{Steinberg:1963wq,Steinberg:2007ih} has proposed the differential-adhesion hypothesis. In this view, cell sorting is due to the interplay of differential adhesion and random cell motility, which progressively replaces weaker intercellular adhesions for stronger adhesion. In addition to differential adhesion, differential surface contraction \cite{HARRIS1976267} aka differential interfacial tension \cite{Brodland:2002gp} due to contraction of the cortical cytoskeleton contribute to the equilibrium configurations of mixed cell aggregates \cite{Krieg:2008jd}. 

Because of its importance for biological development, and the possibility to predict the configuration corresponding with the energy minimum from the differential interfacial energies \cite{Steinberg:1963wq}, cell sorting has become a key benchmark problem for cell-based modeling methodology. Cell sorting has been reproduced in a practically all available cell-based models, including cellular automata \cite{ANTONELLI19731}, vertex-based models \cite{Hutson:2008ec}, center-based models \cite{Graner:1993tg}, and the cellular Potts model \cite{Graner:1992ve,JAGlazier:1993uk},  but small differences are observed \cite{Osborne:2017ff}: The kinetics of cell sorting differs between cell-based modeling methods as well as the extent to which the simulation gets trapped into local minima. Also, methodology relying on single-particles to represent a cell may require unrealistically long interaction lengths or unrealistic cell motility models to achieve complete cell sorting \cite{Osborne:2017ff}

Following previous cellular Potts and vertex-based approaches~\cite{Graner:1992ve,JAGlazier:1993uk,Hutson:2008ec}, we assume that cell motility is governed by volume conservation and an adhesion energy defined at all cell-cell and cell-medium boundaries,
\begin{equation}
H=\lambda_A\sum_{c\in C}\left(A(c)-A_T(c)\right)^2+\sum_{\vec{e}\in E}J\left(\vec{e}\rightarrow L,\vec{e}\rightarrow R\right)\|\vec{e}\|
\label{eq:H2}
\end{equation}
with $A(c)$ and $A_T (c)$ the actual area and resting areas of the cells. The adhesion energy is a sum over all edges $\vec{e}\in E$ in the tissue, with parameter  $J(\vec{e}\rightarrow L,\vec{e}\rightarrow R)$ the adhesion energy per unit interface length, a function of the cells at the left (L) and right (R) sides of the edge.

\subsubsection{Sliding operator enables complete cell sorting}
\label{sec:incomplete}
\begin{figure}
\includegraphics[width=\textwidth]{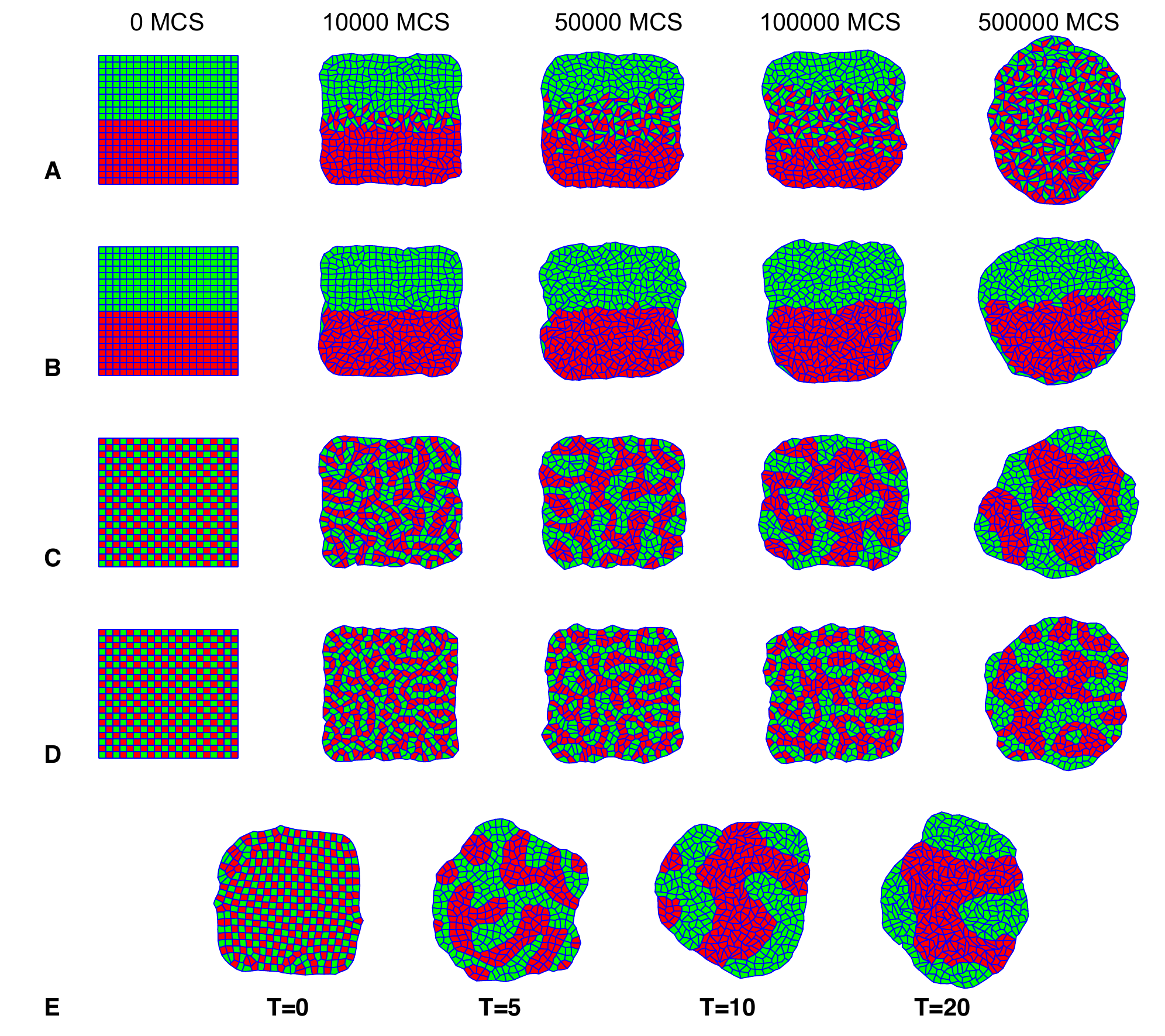}
\caption{{\bf Differential-adhesion driven cell rearrangement in the VirtualLeaf.} Initial condition: 200 green and 200 red cells of $A_T=A(0)=100$, as shown in first column, $l_\mathrm{min}=6$, $l_\mathrm{max}=8$, $\Delta x=0.5$; $T=10$ in panels A-D; (A) Cell mixing ($J(\mathrm{green},\mathrm{green})=J(\mathrm{red},\mathrm{red})=20$, $J(\mathrm{red},\mathrm{green})=10$, $J(\mathrm{cell},\mathrm{medium})=30$), (B) engulfment  ($J(\mathrm{green},\mathrm{green})=20$, $J(\mathrm{red},\mathrm{red})=10$, $J(\mathrm{red},\mathrm{green})=20$, $J(\mathrm{green},\mathrm{medium})=20$, $J(\mathrm{red},\mathrm{medium})=40$); (C) cell sorting ($J(\mathrm{green},\mathrm{green})=20$, $J(\mathrm{red},\mathrm{red})=10$, $J(\mathrm{red},\mathrm{green})=30$, $J(\mathrm{cell},\mathrm{medium})=30$); (D) incomplete cell sorting with only T1 transitions with parameters as in panel C with $\theta_\mathrm{T1}=0.25$; (E) Configurations of cell sorting experiments at 500.000 MCS with increasing values of intrinsic motility (T) with parameters as in panel C. Simulation time is expressed in Monte Carlo Steps (MCS)}
\label{fig:fig4}
\end{figure}

Fig~\ref{fig:fig4}A-C and Videos S1-S3 show the simulation results for three typical settings of the adhesion parameter J. The simulations are initiated with a configuration of $20\times20$ cells of size $10\times10$, with mixed or segregated cell types assignments as shown in the first column of Fig~\ref{fig:fig4}. The target area is set equal to the initial area, at $A_T=100$. The step size for the Monte Carlo algorithm is $\Delta x=0.5$,  $l_\mathrm{min}=6$ and $l_\mathrm{max}=8$. For these parameter settings, the  nodes are moved randomly over a square of side $1/20$ of that of the initial length of the cell-cell interfaces, and one cell-cell interface consists typically of one to two edges, such that sliding moves occur over half to a full cell-cell interface.  

In Fig~\ref{fig:fig4}A the heterotypic adhesion, i.e., the adhesion between green and red cells, is stronger than the homotypic adhesion, i.e., $J(\mathrm{green},\mathrm{green})=J(\mathrm{red},\mathrm{red})>J(\mathrm{red},\mathrm{green})$. The model evolves towards a checkerboard configuration, which maximizes the contact area between red and green cells. Fig~\ref{fig:fig4}B and Fig~\ref{fig:fig4}C show example simulations for which the homotypic adhesion is stronger than the heterotypic adhesion, that is, $J(\mathrm{green},\mathrm{green})=J(\mathrm{red},\mathrm{red})<J(\mathrm{red},\mathrm{green})$.  In addition, in Fig~\ref{fig:fig4}B the  adhesion of the green cells with the surrounding medium is stronger than that of the red cells, i.e., $J(\mathrm{green},\mathrm{ECM})< J(\mathrm{red},\mathrm{ECM})$. Cell sorting requires stochastic boundary movement; at $T=0$ no energetically unfavorable moves are accepted, and the configuration gets stuck at the initial condition, whereas cell sorting is accelerated at higher temperatures  (Figure~\ref{fig:fig4}E). Altogether, in analogy with the cellular Potts model \cite{Graner:1992ve,JAGlazier:1993uk}, the extended VirtualLeaf reproduces the key phenomena related to differential-adhesion driven cell rearrangement: cell sorting, checkerboard pattern formation, and engulfment.

In order to represent cell rearrangements, previous vertex-based simulations applied explicit T1 transitions. A T1 transition rearranges four adjacent cells as shown in Fig~\ref{fig:fig1}B. It is initiated if the length of an intercellular interface,  i.e., an edge $\vec{e}$ connecting a 3-connected node $v_1$ with a second node $v_2$, drops below a threshold, $\|\vec{e}\|<\theta_\mathrm{T1}$. The T1 transition then deletes $\vec{e}$ by fusing $v_1$ and $v_2$ and generates a new edge, $\vec{e}_\perp$, perpendicular to $\vec{e}$. In absence of noise terms, vertex-based models based on such explicit topological transitions generally cannot achieve complete cell sorting, except in specific three-dimensional cases where almost-complete cell sorting can be achieved \cite{Hutson:2008ec}.  

In the extended VirtualLeaf, T1 transitions are represented implicitly as a combination of two sliding moves, where both moves are driven by the Hamiltonian (Figure~\ref{fig:fig1}C). As a first test of the extent to which the sliding operator changes the kinetics of cell sorting, in a second set of simulations we replaced it for explicit T1 transitions.  Fig~\ref{fig:fig4}D and Video S4 show a cell sorting experiment with only T1 transitions and a cellular temperature of $T=10$, the same cellular temperature as that used in Fig.~\ref{fig:fig4}C. Without the sliding operator cell sorting proceeds well over short times, with small clusters of green and red cells forming, but cell sorting remains incomplete. We have currently not investigated the causes of this in detail, but a potential factor is that the sliding operator is fully integrated in the energy minimization processes, in contrast to the traditional treatment of T1 transitions. Also changing the threshold for T1 transitions, currently set at $\theta_\mathrm{T1}=\Delta x/2=0.25$ will likely speed up cell sorting. We will leave a full analysis of the sliding operator relative to the explicit treatment of T1 transitions to the future work. 

\subsubsection{Differential cortical tension}
\begin{figure}
\includegraphics[width=\textwidth]{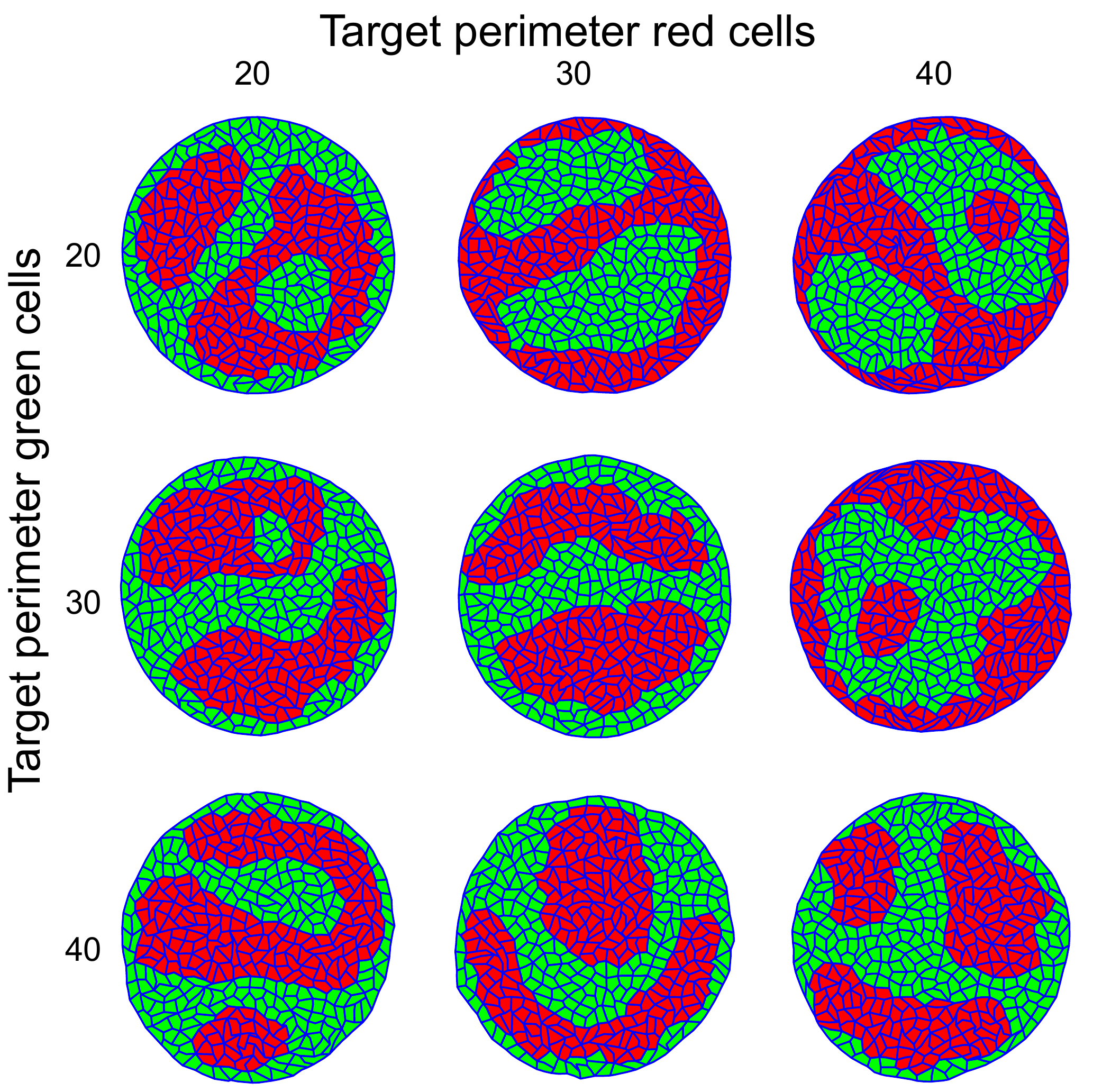}
\caption{{\bf Parameter study of interface specific cortical tension.}
Simulations with cell-type-specific cortical tension applied only at cell medium interfaces, with the target perimeter $P_T(c)$  equal to 20, 30 and 40 as indicated in the axis labels. All other parameters remain unchanged (see Supporting Text S1). This figure shows the tissues after a simulation of 500,000 MCS.}
\label{fig:fig5}
\end{figure}

As an experimental test of the differential-adhesion hypothesis, Krieg and coworkers \cite{Krieg:2008jd} have measured the adhesive forces between induced germline progenitor cells from early zebrafish embryos. The heterotypic adhesion forces between induced endodermal, mesodermal, and ectodermal cells were approximately equal, whereas the homotypic adhesion forces differed between germ layers. Mesodermal cells adhered most strongly to one another, followed by endodermal cells, and ectodermal cells had the weakest adhesive forces to one another. Based on these data, the authors estimated relative values of the adhesion parameters, $J$, in a cellular Potts model. Strikingly, in the Zebrafish germline progenitor aggregates the least coherent ectodermal cells sorted out to the middle of the cellular aggregates. This finding contradicts the differential-adhesion hypothesis (DAH),  which predicts that the least cohesive cells move to the aggregate's periphery, see, e.g., the CPM \cite{Graner:1992ve} and our own simulations (Fig~\ref{fig:fig5}, top-left to bottom-right diagonal). Krieg and coworkers demonstrated that the contradictory prediction can be attributed to differential cortical tension (DCT), an alternative to DAH \cite{HARRIS1976267}, with the highest cortical tension occurring at cell-medium interfaces. To implicitly incorporate cortical tension effects into the Cellular Potts model, Krieg and coworkers reinterpreted the CPM such that a high value of $J$ corresponded with a high interfacial tension. 

To test if VirtualLeaf could represent both DAH and DCT explicitly in the same model framework, we modified the Hamiltonian (Eq.~\ref{eq:H2}) to  add a cell-dependent cortical tension term that is only active at the tissue boundaries. The new Hamiltonian becomes,
\begin{multline}
H=\lambda_A\sum_{c\in C}\left(A(c)-A_T(c)\right)^2+\sum_{\vec{e}\in E}J\left(\vec{e}\rightarrow L,\vec{e}\rightarrow R\right)\|\vec{e}\| + \\
\lambda_\mathrm{cortical}\sum_{\left\{c\in C|c\cap\partial C\right\}}\left(P(c)-P_T(c)\right)^2
\end{multline}
with $\partial C$, the boundary of the tissue, $\lambda_\mathrm{cortical}$, a parameter and $P_T(c)$ a cell type specific target perimeter. $P(c)=\sum_{\vec{e}\in(c\rightarrow E)}\|\vec{e}\|$ is the perimeter of cell $c$, and $P_T (c)$, a target perimeter. Note that the cortical tension term was only applied at the cell-medium interfaces, which would be equivalent to setting $\lambda_\mathrm{cortical}=0$ at cell-cell interfaces. The adhesion parameters were set such that $J(r,r)<J(g,g)<J(g,r)$; i.e., red cells are more coherent than green cells, and red-green interfaces are energetically unfavorable. We have also assumed increased line tension at the boundary of the cell aggregate due to myosin activity \cite{Krieg:2008jd}, by setting $J(l,M)=0$ and $J(d,M)=0$, but this has little effect on the results.

Figure~\ref{fig:fig5} shows a parameter study of this model. If the two cells have equal cortical tension at the boundary of the aggregate (top-left to bottom-right diagonal and Videos S5 and S6) the coherent red cells sort to the center, as expected in absence of additional assumptions. The sorting order is reversed if $P_T(r)>P_T(g)$, thus reducing the cortical tension of red cells relative to that of the green cells (Figure~\ref{fig:fig5} and Video S7). 

\subsection{Epithelial Cell Packing}
The structure of multicellular tissues and the shape of the constituent cells is driven by the interplay of cell division, cell growth, intercellular frictional forces, and global tissue mechanics. Epithelial tissues of plants \cite{Kim:2014dn} and of animals \cite{Farhadifar:2007hr} can be represented by two-dimensional tesselations and are, therefore, a popular model system for studying morphogenesis and emergence of tissue form \cite{Lewis1928}. In particular, the number of neighbors in many epithelial tissues shows a characteristic distribution: Hexagonal cells are the most frequent, followed by pentagonal and heptagonal cells. Although the experimentally observed distribution of can arise due to random cell division alone \cite{MatthewCGibson:2006vp}, the biophysics of cell packing, i.e., programs of cell rearrangement and patterning of interfacial tensions, allows tissues to assume alternative, often narrower (hexagonal) neighbor distributions~\cite{Farhadifar:2007hr}. In absence of cell rearrangements (as e.g., in plant tissues), mathematical simulations have shown that cells must divide over the center of mass and division plane must follow shortest paths, thus forming equally sized, symmetrically shaped daughter cells \cite{Sahlin:2010dd}. 

\subsubsection{VirtualLeaf can reproduce key features of epithelial cell packing}
To see if our model, in particular the flexible cell membranes and the sliding operator, could lead to different predictions for epithelial tissues, we here focus on the results described by Farhadifar et al.~\cite{Farhadifar:2007hr} using the model implementation detailed in Ref.~\cite{Staple:2010bn}. Their vertex-based model uses a Hamiltonian of the form,
\begin{multline}
H=\lambda_A\sum_{c\in C}\left(A(c)-A_T(c)\right)^2+
\sum_{\vec{e}\in E}J \|\vec{e}\|+
\lambda_\mathrm{cortical}\sum_{c\in C}\left(P(c)-P_T(c)\right)^2,
\end{multline}
with $P_T(c)=0$ and $J=J\left(\vec{e}\rightarrow L,\vec{e}\rightarrow R\right)$ the same for all cell interfaces.  In absence of cell division, in this model two distinctive equilibrium cell shape patterns, or ``ground states'' can emerge depending on the parameters. For positive line tension, $J>0$, or negative line tension, $J<0$, with a sufficient high contractility, $\lambda_\mathrm{cortical}$, the global energy minimum in absence of cell divisions (ground state) of the vertex model is a regular, hexagonal tessellation with cellular areas smaller than the target area. The hexagonal tessellation resists compression, expansion or shearing. The alternative global minimum is a `soft network', which occurs at negative line tensions combined with no, or relatively low contractility.  The soft network is characterized by many, alternative irregular tessellations of equal pattern energy, with cellular areas equal to the target area. This soft-to-stiff transition is thought to reflect a soft matter phase transition that accompanies jamming of granular materials  \cite{Atia:2018bp,Tlili:2018gs,Bi:2015hha,Banerjee:ub2018}. 

Farhadifar et al.~\cite{Farhadifar:2007hr} have shown that the cell packing deviates from these global equilibria if cell division is introduced.  The authors picked one cell at random, doubled its target area, and relaxed the cellular configuration to the nearest equilibrium using a conjugate gradient method. They then divided the cell over a randomly oriented axis passing through the cell centroid, after which they relaxed the configuration again to its nearest equilibrium. This procedure was repeated until the tissue consisted of 10,000 cells, after which the topology of the tissue was examined.

\begin{figure}
\includegraphics[width=\textwidth]{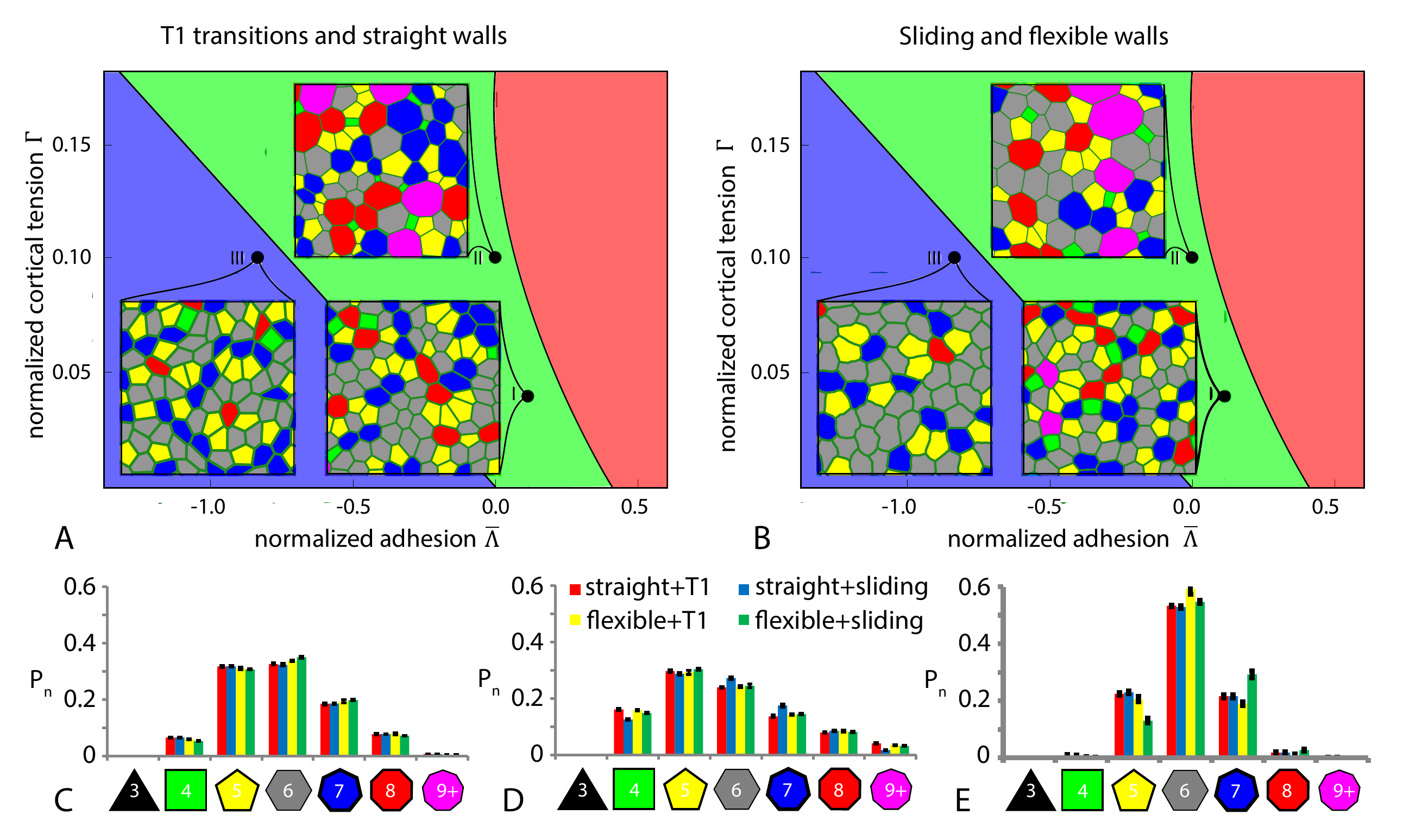}
\caption{{\bf Comparison of straight walls and T1 transitions with flexible walls and sliding on cell morphology.} (A,B) Parametric space as in the article of Farhadifar et al.~\cite{Farhadifar:2007hr}  with three identified morphologies duplicated with VirtualLeaf (case I, II and III) at 100.000 MCS. Vertical axis: normalized cortical tension, $\Gamma = \lambda_\mathrm{cortical}/(\lambda_A A_T)$; horizontal axis: normalized adhesion, $\overline{\Lambda}=J\left(\vec{e}\rightarrow L,\vec{e}\rightarrow R\right)/(\lambda_A A_T^\frac{3}{2})$.  The blue, green, and red regions show the `ground state' of the vertex model in absence of cell division for reference (cf. Figure 1 of Ref.~\cite{Farhadifar:2007hr}): blue, regular hexagonal packing, green: soft networks, red: impossible region. Case I: $\lambda_\mathrm{cortical}=10$, $J\left(\vec{e}\rightarrow L,\vec{e}\rightarrow R\right)=500$. Case II: $\lambda_\mathrm{cortical}=26$, $J\left(\vec{e}\rightarrow L,\vec{e}\rightarrow R\right)=0$. Case III: $\lambda_\mathrm{cortical}=26$, $J\left(\vec{e}\rightarrow L,\vec{e}\rightarrow R\right)=-3560$. Hexagonal networks can be found in the green region of the plot. The sum of cortical tension and adhesion energy is smaller than 0 in the blue region, causing a soft network to occur. Simulations within the red region will be unstable. (C,D,E) Relative amounts of cells with n neighbors when the tissue is in equilibrium. C = case I, D = case II E = case III. The bars represent the averages and the error-bars the standard deviations of 10 time-points between generation 7 and 8. See Supporting Text S1 for detailed simulation descriptions.}
\label{fig:farhadifar}
\end{figure}

To determine if our simulation methods could reproduce these results, we used a vertex-based special case of VirtualLeaf, in which there were only nodes of value 3 and higher, and topological changes occurred through explicit T1 and T2  transitions. We replicated Farhadifar's cell division algorithm with only minor modifications. We picked one cell at random,  slowly increased its target area, and relaxed the tissue to steady state using the Metropolis algorithm. Once the actual area of this cell exceeded twice the target area of the other cells, we let the cell divide over a randomly oriented axis passing through the cell centroid and assigned the original target area to the daughter cells, and the procedure was repeated.  Our simulations (Figure~\ref{fig:farhadifar}A and Videos S8-S10 ) agree visually with the three cases reported previously~\cite{Farhadifar:2007hr} and illustrate the key results of these simulations, displayed upon the ground state diagram by Farhadifar et al.~\cite{Farhadifar:2007hr}. Our vertex-based model replicates a typical ``stiff'' network (Case I), located in the parameter region with a hexagonal ground state, producing cells of approximately uniform size. Furthermore, our model can replicate the outcome of cells with a higher cortical tension (Case II) producing cells with more variable areas and a tessellation that contains large polygons with nine sides or over. Lastly, our model can recapitulate the  `soft network' or ground state (Case III) where cells evolve irregular shapes equal to the target area. 

After eight rounds of cell division, the distribution of polygon classes ($P_n$, the fraction of polygons in the final tissue with $n$ sides) in Case I agree, with only minor differences, with those reported in Ref.~\cite{Farhadifar:2007hr}(red bars in Figure~\ref{fig:farhadifar}C). Both models reveal pentagon and hexagon shaped cells dominate at $P_5\approx0.3$ and $P_6\approx0.3$ while heptagons are slightly less frequent at $P_7\approx0.2$, and tetragons and octogons are present at frequencies of  $P_4\approx P_8\approx0.1$. Our model also produces qualitative agreement in Case II and Case III with those reported for the Vertex Model although our model generated significantly fewer 3, 4, 8 and 9-sided cells. This difference can likely be attributed to the stochasticity in our simulations, which relaxes the configurations more quickly, similar to the effect of annealing reported in Ref.~\cite{Farhadifar:2007hr}.

\subsubsection{Flexible membranes and sliding change Case III, but not Case I and II}
We next tested whether membrane flexibility and the membrane sliding operator could replace algorithm-based T1 transitions to generate a topology indicative of growing tissues. We investigated the performance of these model innovations for three specific cases (\ref{fig:farhadifar}B and Videos S11-S13). For Case I and Case II, the simulations in the presence of sliding and membrane flexibility showed no obvious differences with simulations of the Vertex Model. For Case I (Fig.~\ref{fig:farhadifar}C) and for Case II (Fig.~\ref{fig:farhadifar}D), the distribution of neighbor numbers did not differ between flexible membranes (yellow and green bars) and straight membranes (red and blue bars). Interestingly, for Case III both the visual appearance (Figure~\ref{fig:farhadifar}B) and the neighbor distribution  (Figure~\ref{fig:farhadifar}E) were strongly affected in the presence of sliding and membrane flexibility (green bars): the number of heptagons was higher than for the other simulation conditions, and the number of pentagons was reduced. In absence of membrane flexibility, sliding did not have this effect (blue bars), whereas for membrane flexibility and with T1 transitions, we observed only a small effect (yellow bars). 

In Case I and Case II, the line tension (Case I) or cortical tension (Case II) straightens the cell boundaries, such that boundary flexibility has no effect. In Case III, the specific topology of `soft networks' is due to the boundaries' resistance to compression by adjacent cells. Adding additional nodes to the membranes makes them flexible and allows membranes to buckle (see the `bubbly' boundaries in Figure~\ref{fig:farhadifar}B and Video S13, Case III), which will likely reduce the number of T1 transitions. We did not understand in detail why the effect of neighborhood order is particularly strong in the presence of sliding. A potential explanation is that T1 transitions may introduce spurious energy barriers or time delays between configurations of higher and lower energy, consistent with the incomplete cell sorting discussed in Section~\ref{sec:incomplete}, whereas for sliding such effects are not present. 

\section{Discussion}
In this paper we have introduced extensions of our plant-tissue simulation environment VirtualLeaf~\cite{RMHMerks:2011cj,Merks:2012fu}, adopting it for the simulation of animal tissues. The key novelty is a method to simulate relative movement of cells, the ``sliding operator''. This operator is applied alongside node displacements in a Metropolis-based energy minimization approach.  We have validated the new model using simulations of differential-adhesion driven cell sorting, and found that it can reproduce the key phenomena of differential-adhesion-driven cell sorting, including cell mixing (Figure~\ref{fig:fig4}A), engulfment of one cell type by the other (Figure~\ref{fig:fig4}B), and cell sorting (Figure~\ref{fig:fig4}C). The extended version of VirtualLeaf also reproduces the key phenomena of epithelial cell packing \cite{Farhadifar:2007hr} in 'stiff' regimes of the parameter space, i.e., in case I and II,  where the cell perimeter is under tension ($P(c)>P_T(c)$ and $P_T(c)<2\sqrt(\pi A_T(c))$ (Figure~\ref{fig:farhadifar}B,C,D). In the 'soft' parameter regime (case III), i.e., if the cell perimeter is fully relaxed ($P(c)=P_T(c)$ and $P_T(c)>=2\sqrt(\pi A_T(c))$), the results in VirtualLeaf differ from those reported previously \cite{Farhadifar:2007hr} due to buckling of the cell-cell interfaces.

The sliding operator requires that cell boundaries are represented by multiple nodes, between which the 3-sided vertices can `hop'. Using the sliding operator, topological changes are entirely driven by the energy minimization process through at least two independent moves.  In contrast, in traditional VMs,  T1 transitions are initiated  independently of the energy minimization process, as soon as the length of a cell-cell interface drops below a threshold. This algorithm can artificially introduce T1 'jitters',  in which a configuration repeatedly moves back and forth between two energetically equivalent configurations. Because of this natural integration with the energy minimization algorithm, simulations with the sliding operator more quickly reached complete cell sorting (Figure~\ref{fig:fig4}D) than our simulations with the traditional approach for T1 transitions.  A full quantitative comparison of the two approaches will be left for the future work, and will give more insight into the causes underlying these differences. For example, lowering the interface length threshold $\theta_\mathrm{T1}$ for the T1 transitions will increase their frequency, and will speed up cell sorting for simulations applying only T1 transitions. Similarly, in simulations applying the new slide operator, increasing the probability of slides and movements by increasing the Boltzmann temperature, $T$, increases the speed of cell sorting (Figure~\ref{fig:fig4}E). Preliminary simulations with the new slide operator
suggest that (perhaps somewhat counterintuitively) increasing the step size has little effect on the speed of cell sorting, because it affects only the node moves, but not the node slides that are responsible for cell rearrangement. Interestingly, our preliminary results suggest that increasing the number of nodes used per cell increases the speed of cell rearrangement (Video S14). A possible reason is that in the refined simulations energy barriers are more easily overcome: the smaller moves are associated with a lower, positive $\Delta H$.   A full quantitative analysis of the effect of these and the other parameters on the biological behavior of our model and its computational efficiency will be left to the future work. This analysis will be necessary to decide on the appropriate application domain of this cell-based modeling method among alternative methodology \cite{Osborne:2017ff}. As a first step we plan to perform detailed comparisons with the CPM, which will require quantitative mapping of the model parameters in VirtualLeaf with those of the CPM. Such quantitative parameter mapping are available for the CPM and VMs \cite{Magno:2015bw}, so these will form an excellent starting point.

Several cases of epithelial sheets have been observed that lack straight boundaries between cells including cells in the \textit{Drosophila} amnioserosa \cite{Toyama:2008ga,Solon:2009hg}, cells surrounding the closing blastopore in \textit{Xenopus}~\cite{Feroze:2015hh}, and the jigsaw puzzle cells of the plant epidermis \cite{Fu:2005io,Carter:2017gd,Sapala:elife2018}.  Such irregular boundaries suggests tissue mechanics may be more complex than load-bearing by simple junctional tension \cite{Salbreux:2012bo}. The deviations of cell-cell boundaries from simple lines implies strain on these structures occurs both parallel and perpendicular to the junction and may reflect differential pressures within neighboring cells or the ability of the boundary to bend under compression (e.g., Euler buckling) or  tensions perpendicular to the boundary. The presence of such irregular boundaries becomes more apparent when imaging tissues with higher magnification or when larger cells are sufficiently resolved by lower magnification objectives. The increasing discovery of such irregular shaped cells have precipitated several innovations to VMs and related models~\cite{Fletcher:2017ca}. Bubbly vertex dynamics \cite{Ishimoto:2014hw} 
represents the cell-cell and cell-medium interfaces as curves instead of straight lines, where the curvature is due to  pressure differences. This innovation changes the forces acting upon the vertices, and hence it modifies the dynamics of the VM, but it does not allow buckling of cell boundaries. Buckling is possible thanks to our introduction of multiple cell-cell boundary elements between three cell junctions, which parallels similar innovations in VMs and related methods \cite{Tamulonis:2010kk,Tanaka:2015ku,Perrone:2016fr,Fletcher:2017ca}.  These adaptations allow simulations of tissues that may be under compression in one axis or whose mechanics may be shaped by both medioapical cortical dynamics in addition to junctional contractility. To better simulate such tissues, as a next step we will incorporate a bending stiffness (see, e.g., the bending stiffness of the perimeter in Ref.~\cite{Barton:2017bv}). This will make it possible to explore the full parameter range between the maximally stiff, straight cell-cell interfaces in VMs and the fully floppy cell-cell interfaces that we can currently represent in VirtualLeaf.	

The Hamiltonian and the dynamics of VirtualLeaf \cite{RMHMerks:2011cj} were inspired by both the Cellular Potts model \cite{ArielBalter:2007tf,RMHMerks:2011cj} and by previous Vertex-based Models of plant tissue morphogenesis \cite{Rudge:2005ww,Dupuy:2008ck}. The new sliding operator strengthens the similarity of the VirtualLeaf and the CPM, to the extent that it can be seen as an 'off-lattice' version of the CPM. A related generalization of the CPM was introduced by Scianna and Preziosi \cite{Scianna:2016if}, who have introduced a node-based version of the CPMs. In this generalization the cells can be represented on any tessellation, which has the advantage that the method can be interfaced with a wider range of methods for continuum mechanics where using arbitrary meshes is useful, e.g., for the finite-element method.  The key innovation in their approach, which is shared with VirtualLeaf and VMs, is that cells are represented as polygons. This facilitates simulation of cortical cell tension, but note that efficient, rejection-free implementations of the CPM would typically use very similar data-structures internally. Unlike the VirtualLeaf or the VM, the cell shapes in the node-based CPM are constrained by a tessellation. This possibly introduces similar lattice effects as those found for the CPM, but an advantage of the approach is that it facilitates collision detection. Thus, like the standard CPM, the node-based CPM does not suffer from the limitation of VMs and VirtualLeaf that cell layers must be confluent. Future extensions, e.g.,  new operators for node fusions in conjunction with efficient collision detection algorithms, will relax those limitations of VirtualLeaf. 

In conclusion, with the present extension of a sliding operator, we introduce a new multiparticle method for cell-based modeling and simulation. The method can be categorized within a continuum of closely related multiparticle, Hamiltonian-based methods,  ranging from the CPM \cite{Graner:1992ve,JAGlazier:1993uk}that is run of a regular lattice, via the node-based CPM \cite{Scianna:2016if} that can be run on irregular lattices. VirtualLeaf takes the CPM ``off the lattice'', with the current restriction that tissues must be confluent. Finally, the VM simplifies the representation of the tissue representing them as straight lines \cite{Farhadifar:2007hr}. 


\section{Supplementary Material}

\subsection*{Supplementary Text S1}
Detailed description of the algorithms, the parameter files, and the initial conditions. 

\subsection*{Video S1} 
{\bf Differential-adhesion driven cell rearrangement in VirtualLeaf.} Cell mixing as in Figure~\ref{fig:fig4}A. Initial condition: 200 green and 200 red cells of $A_T=A(0)=100$. $J(\mathrm{green},\mathrm{green})=J(\mathrm{red},\mathrm{red})=20$, $J(\mathrm{red},\mathrm{green})=10$, $J(\mathrm{cell},\mathrm{medium})=30$;   $l_\mathrm{min}=6$, $l_\mathrm{max}=8$, $\Delta x=0.5$; $T=10$. Simulation length: 500,000 Monte Carlo Steps (MCS). View video on \href{https://youtu.be/6uo1fY_He0Q}{YouTube}.

\subsection*{Video S2} 
{\bf Differential-adhesion driven cell rearrangement in VirtualLeaf.} Engulfment as  in Figure~\ref{fig:fig4}B. Initial condition: 200 green and 200 red cells of $A_T=A(0)=100$. $J(\mathrm{green},\mathrm{green})=20$, $J(\mathrm{red},\mathrm{red})=10$, $J(\mathrm{red},\mathrm{green})=20$, $J(\mathrm{green},\mathrm{medium})=20$, $J(\mathrm{red},\mathrm{medium})=40$. Simulation length: 500,000 Monte Carlo Steps (MCS). View video on \href{https://youtu.be/hOdXctm_3CE}{YouTube}.

\subsection*{Video S3} 
{\bf Differential-adhesion driven cell rearrangement in VirtualLeaf.} Cell sorting as in Figure~\ref{fig:fig4}C.  Initial condition: 200 green and 200 red cells of $A_T=A(0)=100$. $J(\mathrm{green},\mathrm{green})=20$, $J(\mathrm{red},\mathrm{red})=10$, $J(\mathrm{red},\mathrm{green})=30$, $J(\mathrm{cell},\mathrm{medium})=30$. Simulation length: 500,000 Monte Carlo Steps (MCS). View video on \href{https://youtu.be/zA0Zx0vwMKk}{YouTube}.

\subsection*{Video S4} 
{\bf Differential-adhesion driven cell rearrangement in VirtualLeaf.} Incomplete cell sorting with only T1 transitions as in Figure~\ref{fig:fig4}D. $\theta_\mathrm{T1}=0.25$; other parameters as in Video S3. Simulation length: 500,000 Monte Carlo Steps (MCS). View video on \href{https://youtu.be/e5XNpfUSS5E}{YouTube}.

\subsection*{Video S5} 
{\bf Effect of interface specific cortical tension.} Simulation with cell-type-specific cortical tension applied only at cell medium interfaces as in Figure~\ref{fig:fig5}, top-left panel. $P_T(\mathrm{red})=20$ at cell-medium interfaces  and $P_T(\mathrm{green})=20$ at cell-medium interfaces. All other parameters have default values (see Supporting Text S1). This figure shows the tissues after a simulation of 500,000 MCS. View video on \href{https://youtu.be/b198iD8BBxk}{YouTube}.

\subsection*{Video S6} 
{\bf Effect of interface specific cortical tension.} Simulation with cell-type-specific cortical tension applied only at cell medium interfaces as in Figure~\ref{fig:fig5}, bottom-right panel. $P_T(\mathrm{red})=40$ at cell-medium interfaces and $P_T(\mathrm{green})=40$ at cell-medium interfaces. All other parameters have default values (see Supporting Text S1). This figure shows the tissues after a simulation of 500,000 MCS. View video on \href{https://youtu.be/CU2QtEExhCk}{YouTube}.

\subsection*{Video S7} 
{\bf Effect of interface specific cortical tension.} Simulation with cell-type-specific cortical tension applied only at cell medium interfaces as in Figure~\ref{fig:fig5}, bottom-right panel. $P_T(\mathrm{red})=40$ at cell-medium interfaces and $P_T(\mathrm{green})=20$ at cell-medium interfaces. All other parameters have default values (see Supporting Text S1). This figure shows the tissues after a simulation of 500,000 MCS. View video on \href{https://youtu.be/mFqNx4fo7Nc}{YouTube}.

\subsection*{Video S8} 
{\bf Simulation of epithelial cell packing} Case I with T1 transitions and straight walls; $\lambda_\mathrm{cortical}=10$, $J\left(\vec{e}\rightarrow L,\vec{e}\rightarrow R\right)=500$. MCS 0 to 40000 with stride 500; cell colors indicate number of neighbors as in Figure~\ref{fig:farhadifar}C-D. View video on \href{https://youtu.be/bE5FTZQue3Q}{YouTube}.

\subsection*{Video S9} 
{\bf Simulation of epithelial cell packing} Case II with T1 transitions and straight walls; $\lambda_\mathrm{cortical}=26$, $J\left(\vec{e}\rightarrow L,\vec{e}\rightarrow R\right)=0$. MCS 0 to 40000 with stride 500; cell colors indicate number of neighbors as in Figure~\ref{fig:farhadifar}C-D. View video on \href{https://youtu.be/XMR6E_1krOQ}{YouTube}.

\subsection*{Video S10} 
{\bf Simulation of epithelial cell packing}  Case III with T1 transitions and straight walls;  $\lambda_\mathrm{cortical}=26$, $J\left(\vec{e}\rightarrow L,\vec{e}\rightarrow R\right)=-3560$. MCS 0 to 40000 with stride 500; cell colors indicate number of neighbors as in Figure~\ref{fig:farhadifar}C-D. View video on \href{https://youtu.be/QTwRHiddnTY}{YouTube}.

\subsection*{Video S11} 
{\bf Simulation of epithelial cell packing} Case I with sliding and flexible walls; $\lambda_\mathrm{cortical}=10$, $J\left(\vec{e}\rightarrow L,\vec{e}\rightarrow R\right)=500$. MCS 0 to 40000 with stride 500; cell colors indicate number of neighbors as in Figure~\ref{fig:farhadifar}C-D. View video on \href{https://youtu.be/097eRzyQ7gY}{YouTube}.

\subsection*{Video S12} 
{\bf Simulation of epithelial cell packing} Case II with sliding and flexible walls; $\lambda_\mathrm{cortical}=26$, $J\left(\vec{e}\rightarrow L,\vec{e}\rightarrow R\right)=0$. MCS 0 to 40000 with stride 500; cell colors indicate number of neighbors as in Figure~\ref{fig:farhadifar}C-D. View video on \href{https://youtu.be/7MnJsU5crjs}{YouTube}.

\subsection*{Video S13} 
{\bf Simulation of epithelial cell packing}  Case III with sliding and flexible walls;  $\lambda_\mathrm{cortical}=26$, $J\left(\vec{e}\rightarrow L,\vec{e}\rightarrow R\right)=-3560$. MCS 0 to 40000 with stride 500; cell colors indicate number of neighbors as in Figure~\ref{fig:farhadifar}C-D. View video on \href{https://youtu.be/1O5klfORzEs}{YouTube}.

\subsection*{Video S14}
{\bf Effect of cell resolution on cell sorting kinetics.} Left, control simulation of cell mixing (cf. Figure~\ref{fig:fig4}A) with default values of $l_\mathrm{min}=6$ and $l_\mathrm{max}=8$; Right, refined simulation of cell mixing with reduced values of $l_\mathrm{min}=3$ and $l_\mathrm{max}=4$ such that twice the number of edges and nodes is used for each cell. Bottom panel shows the summed length of red-green cell-cell interfaces relative to the total length of all cell-cell interfaces in the configuration,
\begin{equation}
\frac{1}{\sum_{\vec{e}\in E}\|\vec{e}\|}%
\sum_{\left\{\vec{e}\in E | %
\vec{e}\text{ is red-green interface}\right\}}\|\vec{e}\|%
,
\end{equation}
as a function of time. 
The moving dot indicates the present time. 
View video on \href{https://youtu.be/sQ_7LpVDvVk}{YouTube}.




\section*{Supplementary material}

**Supplementary information to
Wolff, Davidson and Merks, Adapting a plant tissue model to animal development:
introducing cell sliding into *Virtual Leaf***

1. **Parameter Settings**

**1.1. DAH Simulations**
Initial settings: 200 green (g) cells and 200 red (r) cells (Figure S1A and B).
Cells move by sliding. $T = 10$, $\lambda_A = 1$, $A_T = A_{t=0}$, $l_{min} = 6$, $l_{max} = 8$.

General Parameters:
Cell sorting: $\lambda_P = 0$, $\lambda_{l(g,g)} = 20$, $\lambda_{l(r,r)} = 10$, $\lambda_{l(g,r)} = 30$, $\lambda_{l(g,M)} = 30$, $\lambda_{l(r,M)} = 30$.
Checkerboard: $\lambda_P = 0$, $\lambda_{l(g,g)} = 20$, $\lambda_{l(r,r)} = 20$, $\lambda_{l(g,r)} = 10$, $\lambda_{l(g,M)} = 30$, $\lambda_{l(r,M)} = 30$.
Engulfment: $\lambda_P = 0$, $\lambda_{l(g,g)} = 20$, $\lambda_{l(r,r)} = 10$, $\lambda_{l(g,r)} = 20$, $\lambda_{l(g,M)} = 20$, $\lambda_{l(r,M)} = 40$.
Stochasticity: $\lambda_P = 0$, $\lambda_{l(g,g)} = 20$, $\lambda_{l(r,r)} = 10$, $\lambda_{l(g,r)} = 30$, $\lambda_{l(g,M)} = 30$, $\lambda_{l(r,M)} = 30$.
Cortical tension: $\lambda_P = 20$, $\lambda_{l(g,g)} = 20$, $\lambda_{l(r,r)} = 10$, $\lambda_{l(g,r)} = 30$, $\lambda_{l(g,M)} = 0$, $\lambda_{l(r,M)} = 0$.

Experiment specific parameters:
Stochasticity experiment: $T = 0$, $T = 5$, $T = 10$, $T = 20$
Cortical tension: $P_{t\,(red)} = 20$ $P_{t\,(green)} = 20$, $P_{t\,(red)} = 20$ $P_{t\,(green)} = 30$, $P_{t\,(red)} = 20$ $P_{t\,(green)} = 40$,
$P_{t\,(red)} = 30$ $P_{t\,(green)} = 20$, $P_{t\,(red)} = 30$ $P_{t\,(green)} = 30$, $P_{t\,(red)} = 30$ $P_{t\,(green)} = 40$,
$P_{t\,(red)} = 40$ $P_{t\,(green)} = 40$, $P_{t\,(red)} = 40$ $P_{t\,(green)} = 30$, $P_{t\,(red)} = 40$ $P_{t\,(green)} = 40$,

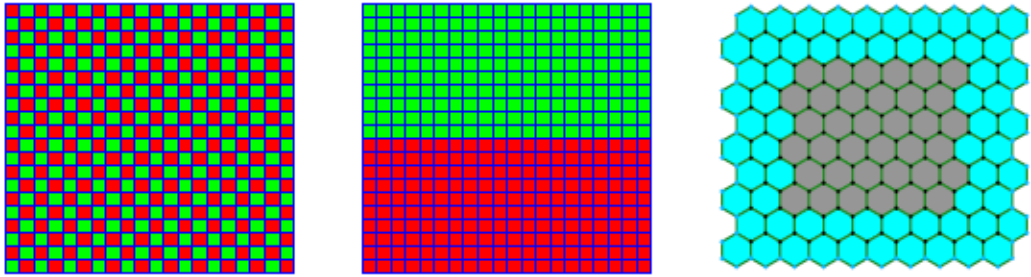

*Figure S1 Initial configurations of simulations. A) Initial configuration for: cell sorting, stochasticity and cortical tension simulations. B) Initial configuration for: checkerboard and engulfment simulations. C) Initial configuration for: Farhadifar simulations.*

**1.2. Farhadifar Simulations**
Initial conditions for all Farhadifar simulations:
Simulations start with 36 hexagonal cells (grey) + 28 hexagonal boundary cells (cyan)
(Figure S1C), and $T = 10$, $A_T = 260$, $P_T = 0$.

Parameters that define case I, II and III in the paper of Farhadifar et al. were used.
Thus, $\lambda_A$ is a relative parameter and is set to 1 and $\lambda_P$ and $\lambda_l$ are defined as:
$\lambda_P = \bar{\Gamma}\lambda_A A_T$ and $\lambda_l = \bar{\Lambda}\lambda_A A_T^{3/2}$. Initial areas of all cells $A_0$ (and resulting perimeters $P_0$) were chosen so that all forces in the tissue start in equilibrium. Furthermore,
$l_{max} = P_0/12$ and $l_{min} = P_0/24$. The following parameters result from this:

Case I: $\lambda_A = 1$, $\lambda_P = 10$, $\lambda_l = 500$, $A_0 = 134$, $l_{max} = 3.59$, $l_{min} = 1.80$.
Case II: $\lambda_A = 1$, $\lambda_P = 26$, $\lambda_l = 0$, $A_0 = 53$, $l_{max} = 2.26$, $l_{min} = 1.13$.
Case III: $\lambda_A = 1$, $\lambda_P = 26$, $\lambda_l = -3560$, $A_0 = 260$, $l_{max} = 5.00$, $l_{min} = 2.50$.

Simulations end when generation eight is reached. A generation is defined as $\log_2(N_c/N_0)$, with $N_c$ as the number of cells, and $N_0$ as the initial amount of cells (36). The time-points used for the histograms in figure 6 are the last ten generation time-points (7.1, 7.2, 7.3, 7.4, 7.5, 7.6, 7.7, 7.8, 7.9 and 8.0). The bars indicate the averages of these time-points and the error-bars indicate their standard-deviations.

The Farhadifar simulations use a cell division algorithm in which one random cell is selected to grow. After the cell has doubled in size it divides and another random cell is selected.

The same colors have been used in all the figures as by Farhadifar et al.: black 3 neighbors, green, 4 neighbors, yellow 5 neighbors, grey 6 neighbors, blue 7 neighbors, red 8 neighbors, magenta 9 neighbors, cyan boundary cells.

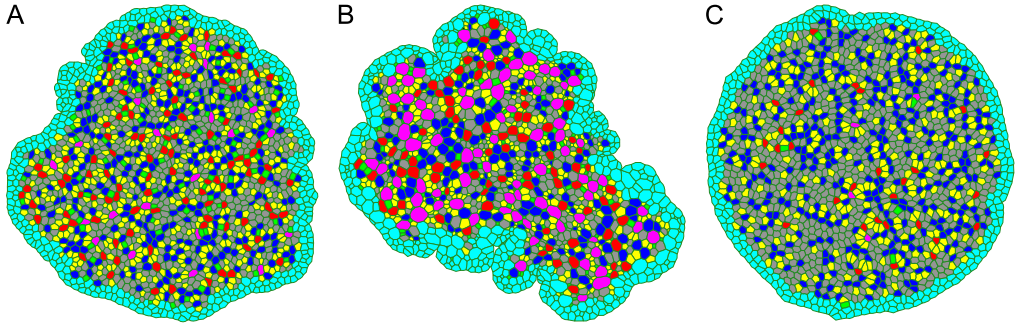

*Figure S2 Tissue shapes of simulations with sliding and flexible walls at 1.000.000 MCS. Differences in tissue shape are caused by $\lambda_P$ and $\lambda_l$. (A) Case I (B) Case II (C) Case III.*

2. **Sliding Algorithm**

---

**Algorithm 1** Sliding Algorithm
---
**In:** random vertex *n*
**If:** vertex = 3-vertex **then**
    Select random slide direction
    **If:** cell is concave $(x_n - x_{n-1})(y_{n+1} - y_n) < (x_{n+1} - x_n)(y_n - y_{n-1})$ **then**
        reject slide **return** 0
    **For:** all edges of cells connected to n
        **if:** new edge intersects existing edge **then**
            reject slide **return** 0
    **Until:** all edges have been tested
    Calculate $\Delta H$ for slide
    **If:** $\Delta H < -threshold$ **or** $\mathrm{RANDOM} < e^{(-\Delta H - threshold)/T}$ **then**
        accept slide, update configuration
    **Else:** $\Delta H = 0$
**Return:** $\Delta H$


\end{document}